\documentclass[aps,prl,twocolumn,showpacs,preprintnumbers,amsmath,amssymb,floats]{revtex4}

\usepackage{graphicx}
\usepackage{epsfig}
\usepackage{latexsym}
\usepackage{dcolumn}
\usepackage{bm}
\usepackage{amsfonts}

\begin{document}
\preprint{APS/preprint}

\title{Electronic Charge Order in the Pseudogap State of Bi$_{2}$Sr$_{2}$CaCu$_{2}$O$_{8+\delta}$}

\author{Y. H. Liu, K. Takeyama, T. Kurosawa, N. Momono, M. Oda and M. Ido}
\affiliation{Department of Physics, Hokkaido University, Sapporo
060-0810, Japan}

\begin{abstract}
Scanning tunneling microscopy/spectroscopy is used to examine
the $4a$$\times$$4a$ electronic charge order (CO) in the pseudogap (PG)
state above {\it T}$_{\rm c}$ on Bi$_{2}$Sr$_{2}$CaCu$_{2}$O$_{8+\delta}$.
It is demonstrated that the static CO develops markedly in the inhomogeneous PG state,
while it is very weak in the homogeneous PG state.  We suggest that this static CO,
which is considered to be stabilized by the pinning of the dynamically fluctuating CO,
will remain below {\it T}$_{\rm c}$, together with the inhomogeneous gap structure,
and coexist with the superconductivity.

\end{abstract}


\maketitle

In high-{\it T}$_{\rm c}$ cuprate superconductors, it has been established that an unusual
electronic state, characterized by a gap-like structure around the Fermi level ${\it E}_{\rm F}$, the so-called ``pseudogap (PG)'',
develops in the normal state above {\it T}$_{\rm c}$, and it must be
well understood to elucidate the mechanism of high-{\it T}$_{\rm
c}$ superconductivity. Recently, in the PG state of underdoped (UD) Bi$_{2}$Sr$_{2}$CaCu$_{2}$O$_{8+\delta}$
(Bi2212), Vershinin {\it et al.} \cite{rf:Vershinin} found a charge
order (CO) in two-dimensional (2D) maps of the local density of states (LDOS)
at specified energies, which were obtained by scanning tunneling microscopy/spectroscopy (STM/STS).
This CO is oriented along the two Cu-O bond directions,
intersecting at right angles; its period is independent of energy, $4.5a$$\sim$$4.8a$ along each
Cu-O direction ($a$: the lattice constant or the Cu-O-Cu distance),
which is called ``nondispersive.'' Interestingly, the nondispersive CO develops markedly at low energies
within the PG and tends to disappear outside the PG. On the other hand, in the SC state, they did not
observe the nondispersive CO, but observed dispersive LDOS modulations due to
quasiparticle interference effects, which were first reported by Hoffman {\it et al.} \cite{rf:Hoffman}.
This tempted many researchers to suppose that the nondispersive CO is a characteristic feature only for the PG state above {\it T}$_{\rm c}$.

However, in the SC state of UD Bi2212, Howald {\it et al.}
\cite{rf:Howald} and Momono {\it et al.} \cite{rf:Momono}
observed a nondispersive CO, whose period, $\sim$$4a$ for each Cu-O direction, was smaller than that
reported by Vershinin {\it et al.} in the PG state. Furthermore, it
has recently been demonstrated in STM/STS experiments by Hashimoto
{\it et al.} \cite{rf:Hashimoto} that the amplitude of
$4a$$\times$$4a$ CO at {\it T}$\ll${\it T}$_{\rm c}$ is
strongly sample-dependent; in samples exhibiting an intense
$4a$$\times$$4a$ CO at {\it T}$\ll${\it T}$_{\rm c}$, the
spatial dependence of the energy gap structure is inhomogeneous
on the nanometer scale, and vice versa. Whether the $4a$$\times$$4a$
CO is a common feature for both the SC and PG states is of great
interest for elucidating the relation among the CO, PG and
high-${\it T}_{\rm c}$ superconductivity. In this paper, we report STM/STS
experiments in the PG state above {\it T}$_{\rm c}$ on two kinds of samples that exhibit strong and
weak $4a$$\times$$4a$ CO's at {\it T}$\ll${\it T}$_{\rm c}$, as shown in Figs. 2 (d) and (e),
and suggest that the static $4a$$\times$$4a$ CO, which develops markedly in the inhomogeneous PG state,
will remain below ${\it T}_{\rm c}$, together with the inhomogeneous gap structure, and coexist with the superconductivity.

Bi2212 crystals were grown by the traveling solvent floating zone method. The {\it
T}$_{\rm c}$ was determined to be 81 K from the SC diamagnetic transition curve, and the
hole doping level ${\it p}$ was estimated to be slightly UD, $\sim$0.14. The details of
this estimation have been reported in Ref. 7. In this study, two samples, labeled L and
M, were cleaved in a vacuum of $<$$10^{-9}$ torr at liquid-nitrogen
temperature and room temperature, respectively, before being inserted in situ into an
STM unit at $\sim$7 K. It should be noted that STM/STS experiments in the PG state were
performed after finishing those at {\it T}$\ll${\it T}$_{\rm c}$ and warming the sample
gradually in situ up to a temperature above {\it T}$_{\rm c}$. STM images were taken in
the constant-height mode; under the constant sample-bias voltage ${\it V}_{\rm s}$, the
tip-height, determined by giving the initial value ${\it I}_{\rm t}^{0}$ for tunneling
current {\it I}$_{\rm t}$ at the initial tip-position, was kept constant during the tip
scanning, where {\it I}$_{\rm t}$ was measured as a function of tip-position.

As is well known, the cleavage in Bi2212 usually occurs between the semiconducting Bi-O
planes with an energy gap ${\it E}_{\rm g}$ of the order of 100 meV, forming a bilayer,
in which excess oxygen ions, providing the Cu-O plane nearby with mobile holes, are contained.
In STM experiments on the cleaved surface, the top most atomic plane closest to the tip is the
semiconducting Bi-O plane, the second the insulating Sr-O plane and the
third the metallic or SC Cu-O plane, and we can observe the
different planes selectively, as reported in Refs. 6 and 8; STM images measured at ${\it
V}_{\rm s}$$>$${\it E}_{\rm g}/{\it e}$ $\sim100$~mV reflect the Bi-O plane,
while STM images measured at $V_{\rm s}$\raisebox{-0.1ex}{$\stackrel{<}{_{\sim}}$}100 mV mainly
reflect the Cu-O plane. Here, we focus on low-bias STM images reflecting the Cu-O plane.

\begin{figure}[t]
\begin{center}
\includegraphics[scale=.43]{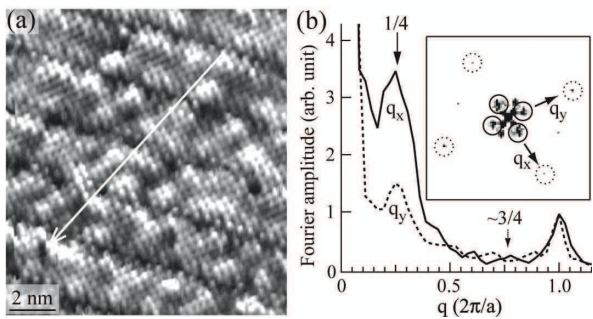}
\vspace*{-0.15cm}
\caption{(a) STM image for an area of sample L, obtained at $T=85$~K, ${\it
V}_{\rm s}=30$~mV and $I_{\rm t}^{0}=0.18$ nA. (b) Line cuts taken along the ${\it q}_{x}$
and ${\it q}_{y}$ directions on the Fourier map (inset).}\label{figure1}
\end{center}
\end{figure}

\begin{figure}[t]
\begin{center}
\includegraphics[scale=.52]{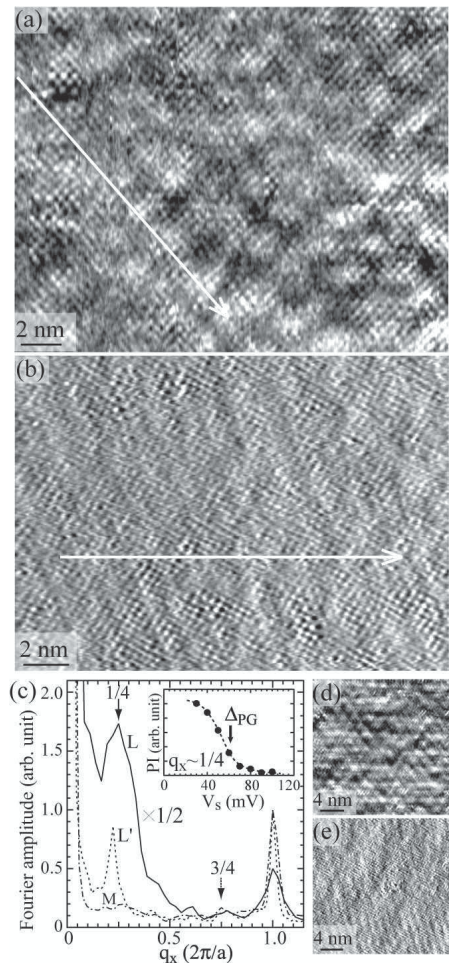}
\vspace*{-0.08cm} \caption{(a) STM image for an another area (L') of sample L obtained at $T=85$~K, ${\it V}_{\rm s}=30$~mV
and $I_{\rm t}^{0}=0.18$~nA. (b) STM image for sample M obtained at $T=88$~K, ${\it V}_{\rm s}=20$~mV and $I_{\rm t}^{0}=0.18$~nA.
(c) Line cuts taken along the ${\it q}_{x}$ direction on the Fourier maps corresponding to the two areas of sample L and sample M.
Inset: ${\it V}_{\rm s}$-dependence of the peak intensity (PI) at ${\it q}_{x}$$\sim$$2\pi/4a$ measured in the latter area (L') of sample L.
The PI reaches the background level above ${\it V}_{\rm s}$$\sim$$90$~mV. (d) and (e) STM images for samples L and M obtained at $T\sim$7 K in
the SC state, respectively (${\it V}_{\rm s}=30$~mV).}\label{figure2}
\end{center}
\end{figure}

Shown in Fig. 1 (a) is an STM image measured at {\it V}$_{\rm s} = 30$
mV on sample L in the PG state (${\it T}=85$~K) above ${\it T}_{\rm c}$. One can see a
2D charge modulation or a 2D CO clearly. It is locally distorted but oriented along the Cu-O directions tilted by
$45^{\circ}$ from a weak 1D superstructure. The 1D superstructure
is observed much more evidently at high voltages of several hundred mV
($>$${\it E}_{\rm g}/{\it e}$), because it is a striking feature of the Bi-O plane \cite{rf:Oda2}.
To quantify the period of 2D CO, we performed a Fourier transform on the image.
The Fourier map is shown in the inset of Fig. 1 (b), where four
spots surrounded by dotted-line circles correspond to the atomic periodicity and four broad
spots surrounded by solid-line circles near the origin result from the periodic CO.
The most intense peaks appearing at $(2\pi/4a, 0)$ and $(0, 2\pi/4a)$ in the line cuts
(Fig. 1 (b)) taken along the ${\it q}_{x}$ and ${\it q}_{y}$
directions on the Fourier map indicate that the 2D CO has an average period of $4a$$\times$$4a$.
Very small peaks at around $(6\pi/4a, 0)$ and $(0, 6\pi/4a)$ seem to correspond to the internal
structure of the CO, which has been observed clearly in the
low-temperature PG state of lightly-doped Ca$_{2-x}$Na$_{x}$CuO$_{2}$Cl$_{2}$ \cite{rf:Hanagury}.

Figure 2 (a) shows an STM image measured at {\it V}$_{\rm s}=30$~mV in another area of
sample L, $\sim$60 nm from the former area where the STM image of Fig. 1 (a) was
taken. A similar CO can be seen throughout this image. The Fourier
analysis verified that the CO in Fig. 2 (a) had a period of
$4.4a$$\times$$4.4a$, which was slightly larger than that in Fig. 1 (a). STM measurements in
several other areas, where a CO can be clearly seen in low-bias images with an atomic resolution,
demonstrated that the period of the CO, distributed slightly in the range from $4a$ to $4.4a$,
was smaller than the value ($4.5a$$\sim$$4.8a$) reported by Vershinin {\it et al} in the PG state \cite{rf:Vershinin},
but almost the same as in the SC state of sample L. Hereafter, we denote the period near four times the
lattice-constant as $\sim$$4a$$\times$$4a$.

From the {\it V}$_{\rm s}$-dependence of the STM image in the latter area of sample L,
it was found that the period of the CO was energy-independent, {\it i.e.} nondispersive, while
its amplitude decreased rapidly with increasing energy and became negligibly small above the PG energy,
$\sim$$\Delta_{\rm PG}$, as shown in the inset of Fig. 2 (c). This, consistent with the
result observed by Vershinin {\it et al.}~\cite{rf:Vershinin}, indicates that the
characteristic energy of the $\sim$$4a$$\times$$4a$ charge order in the PG state above
$T_{\rm c}$ is the corresponding energy gap, as in the SC state below $T_{\rm c}$
\cite{rf:Momono,rf:Hashimoto}.

Figure 2 (b) shows an STM image in the PG state of sample M, measured at $T=88$ K  and $V_{\rm s}=20$ mV,
which was the lowest of the examined bias voltages. For this sample, similar images were obtained at $V_{\rm
s}$\raisebox{-0.1ex}{$\stackrel{<}{_{\sim}}$}100 mV. In the low-bias image of sample M,
one can see that an atomic resolution is obtained as in those of sample L, while the
CO is too weak to be distinguished; indeed, the peak at ($\sim$$2\pi/4a, 0$) in
the Fourier map (Fig. 2 (c)), corresponding to the $\sim$$4a$$\times$$4a$ CO, is
rather small compared with the Bragg peak at $(2\pi/a, 0)$. The former fact guarantees
that the obtained image is reliable, and it is therefore concluded that such a weak CO is an intrinsic
property in the PG state of sample M. Thus, the present two samples, L and M, exhibit strong and weak CO's
in the PG state, respectively. It should be remembered here that samples L and M also exhibit strong and
weak CO's (Figs. 2 (d) and (e)) at {\it T}$\ll${\it T}$_{\rm c}$, respectively. These facts tempt us to
suppose that if the $\sim$$4a$$\times$$4a$ CO develops markedly in the PG state above $T_{\rm c}$,
it will continue to exist in the SC state below $T_{\rm c}$.

\begin{figure}[t]
\begin{center}
\includegraphics[scale=.40]{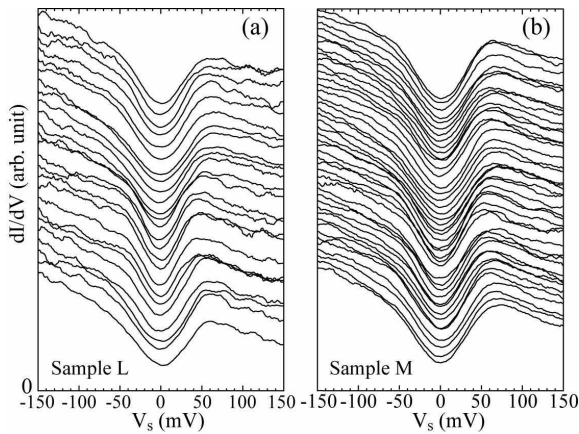}
\vspace{-0.2cm} \caption{STS spectra measured along the white lines
in the images of Fig. 2. The labels (a) and (b) in Fig. 3 correspond to those in Fig. 2.
The spectra are normalized with the value at {\it V}$_{\rm s}=-150$~mV and shifted along the
vertical axis for clarity. The zero level in the vertical axis is for the lowest spectrum.}\label{figure3}
\end{center}
\end{figure}

As reported in Ref. 6, in samples exhibiting the strong CO at
{\it T}$\ll${\it T}$_{\rm c}$, the spatial dependence of the energy gap
structure is inhomogeneous on the nanometer scale, and vice versa.
We will demonstrate in the following paragraphs that such a
correlation between the CO and the gap inhomogeneity also
holds in the PG state above {\it T}$_{\rm c}$, which strongly supports the
contention that the CO in the SC state will be the same as that in the PG state.

Figures 3 (a) and (b) are tunneling conductance d$I$/d$V$-$V_{\rm s}$ curves, referred to as STS spectra, which were measured along
the solid lines in Figs. 2 (a) and (b), STM images of samples L
and M, respectively. Similar PG structures are seen in the STS spectra,
regardless of the sample and position. The d$I$/d$V$ value, which tends
to increase gradually with the lowering of $V_{\rm s}$, is largely reduced
in the range around $V_{\rm s}=0$, corresponding to $E_{\rm F}$; thus, it exhibits a broad peak around the positive
voltage $V_{\rm s}^{\rm p}$, while a broad bend appears around $-V_{\rm s}^{\rm p}$ in the
negative $V_{\rm s}$-region. Here, we define the energy size of PG,
$\Delta_{\rm PG}$, from the peak position, $V_{\rm s}^{\rm p}$; $\Delta_{\rm PG}{\equiv}eV_{\rm s}^{\rm p}$.

Comparing the two sets of STS spectra in Fig. 3, it is found that the $\Delta_{\rm PG}$ value changes as a function of position for sample L (Fig. 2 (a)), which exhibits a strong CO,
while it is more homogeneous for sample M (Fig. 2 (b)), which exhibits a very weak CO;
the average PG values for the two samples are almost the same ($\sim$$60$ meV), meaning that they have little difference
in $p$, but the standard deviation (7.0 meV) for sample L is $\sim$3 times larger than that (2.5 meV)
for sample M.  Such a difference in the spatial inhomogeneity of $\Delta_{\rm PG}$ between the two samples can be seen
more clearly in 3D illustrations (Figs. 4 (b) and (c)) of the STS spectra (Figs. 3 (a) and (b)),
where the vertical axis is for tunneling conductance, and the horizontal axes are for bias voltage and position, respectively.

\begin{figure*}[t]
\begin{center}
\includegraphics[scale=.35]{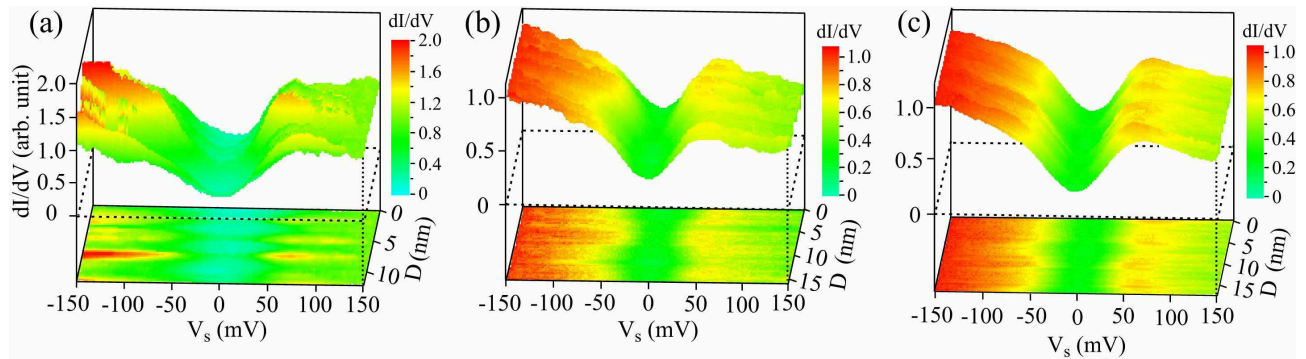}
\caption{3D illustrations of STS spectra, measured
along the white lines in Fig. 1 (a), Figs. 2 (a) and (b), respectively.
One of the horizontal axes is for the distance ($D$) from the initial measurement position.
Each spectrum in (b) and (c) was normalized with its value at {\it V}$_{\rm s}$=-150~mV, as in Fig. 3,
while each spectrum in (a) was normalized with its value at {\it V}$_{\rm s}=150$~mV
because one of the spectra involves a big noise at {\it V}$_{\rm s}=-150$~mV.}\label{figure4}
\end{center}
\end{figure*}

A 3D illustration of STS spectra measured along a straight line in the former area (Fig. 1 (a)) of sample L,
which exhibits an intense CO, is shown in Fig. 4 (a), as well. The average $\Delta_{\rm PG}$ value for this area, is $\sim$90 meV,
suggesting that the $p$ value is smaller than that of the other area (Fig. 2 (a)). Since the $\sim$$4a$$\times$$4a$ CO instability
tends to be enhanced with the lowering of $p$, the reduction of $p$ will be one of the factors that cause the intense CO \cite{rf:Momono,rf:Hashimoto}.
Furthermore, combining the STM image (Fig. 1 (a)) and the corresponding STS spectra (Fig. 4 (a)), one can see that in the area exhibiting the
intense CO, the degree of suppression of spectral weights around $V_{\rm s}=0$ ($E_{\rm F}$), as well as the $\Delta_{\rm PG}$ value,
changes drastically as a function of position, that is, the PG structure is much more inhomogeneous. Thus, the energy gap inhomogeneity
correlates strongly with the development of $\sim$$4a$$\times$$4a$ CO in the PG state, as in the SC state.

To understand such a correlation between the electronic CO and the gap inhomogeneity,
we have proposed that the CO is dynamic in itself and, if Bi2212 samples involve strong scattering centers for quasiparticles,
leading to the gap inhomogeneity, such as crystallographic imperfections, the scattering centers will function as effective
pinning centers for the dynamically fluctuating CO and make it static \cite{rf:Hashimoto}.
On the basis of this pinning picture, the dynamically fluctuating CO would be a candidate for the hidden order
of the homogeneous PG state and, the degree of development of the static CO, especially in the samples with similar
doping levels, can be explained in terms of the density and/or strength of pinning centers.

According to ARPES experiments on UD Bi2212 \cite{rf:Norman}, the PG starts to develop around
temperature ${\it T}^{\rm *}$, well above {\it T}$_{\rm c}$, on a part of the Fermi surface (FS)
near $(\pi/a, 0)$ and evolves gradually toward the node point of the $d$-wave gap near ($\pi$/a, $\pi$/a)
with the lowering of $T$, but an ungapped region still remains around the node point just above {\it T}$_{\rm c}$,
leading to an arc-shaped FS, which is referred to as the Fermi arc (FA). The FS parts, inside and outside the FA,
have been considered to consist of coherent and incoherent electronic states, respectively \cite{rf:Pines,rf:Geshkenbein,rf:Furukawa,rf:Wen,rf:Yanase}.
In light of these facts, we have argued that even if incoherent quasiparticles outside the FA form pairs
in the PG state below $\sim$${\it T}^{\rm *}$, they cannot establish long-range phase coherence in collective motion,
which will be done by the pairing of coherent quasiparticles on the FA, and the energy gap which opens up on the
FA below {\it T}$_{\rm c}$ will function as an effective SC gap in determining {\it T}$_{\rm c}$ \cite{rf:Oda3}.
It should be remembered here that the PG, which is formed on the incoherent part of the FS, is spatially inhomogeneous
in samples exhibiting the strong, pinned $\sim$$4a$$\times$$4a$ CO, and vice versa.  This fact is naturally understandable,
because incoherent electronic states are easily modified by external perturbation, which is due to the randomness associated
with pinning potentials for the CO.  Furthermore, since the $\sim$$4a$$\times$$4a$ CO can be seen in almost the same energy (bias-voltage)
range as the PG, incoherent, antinodal quasiparticle or pair states outside the FA will also be responsible for the CO.
In fact, it has been found in STM/STS experiments at {\it T}$\ll${\it T}$_{\rm c}$ that at low energies around {\it E}$_{\rm F}$,
reflecting the quasiparticle states inside the FA, the gap structure is characterized by a spatially-homogeneous $d$-wave gap and
the CO tends to fade out, while at high energies around the gap edge, reflecting the quasiparticle states outside the FA, the gap
structure is strongly inhomogeneous and the CO becomes marked \cite{rf:Hashimoto}.  Thus, it is suggested that if the $\sim$$4a$$\times$$4a$ CO,
which is considered to be dynamic in itself and associated with antinodal quasiparticle or pair states outside the FA, is pinned down and static
in the inhomogeneous PG state above {\it T}$_{\rm c}$, it will remain below {\it T}$_{\rm c}$, together with the inhomogeneous gap structure in
the antinodal region, and coexist with the superconductivity caused by the pairing of coherent quasiparticles on the FA, that is,
the so-called ``FA superconductivity" \cite{rf:Hashimoto,rf:Oda3,rf:McElroy}.

In summary, we performed STM/STS experiments in the PG state above {\it T}$_{\rm c}$ on UD Bi2212,
and demonstrated that a strong correlation between the static $\sim$$4a$$\times$$4a$ CO and the gap
inhomogeneity held in the PG state as in the SC state; the static CO develops markedly in the inhomogeneous PG state,
whereas it is very weak in the homogeneous PG state.  The static CO can be understood in terms of a pinning picture in
which the dynamically fluctuating CO, which seems to be intrinsic in the homogeneous PG and SC states, is pinned down by
scattering potentials, leading to gap inhomogeneity.  We also suggested that the $\sim$$4a$$\times$$4a$ CO is associated with
incoherent quasiparticle or pair states on the antinodal FS, which are responsible for the PG,  and can coexist with the FA
superconductivity below {\it T}$_{\rm c}$.  At the present stage, it is unclear whether a Cooper-pair's CO or a bosonic CO
is realized in high-{\it T}$_{\rm c}$ cuprates. However, the present study will be useful for discussions about the propriety
of theoretical models on the $\sim$$4a$$\times$$4a$ CO \cite{rf:Kivelson,rf:Bosch,rf:Podolsky,rf:Tesanovic,rf:Ohkawa}.

The authors thank Prof. F. J. Ohkawa for valuable discussions. This work was
supported by the $21^{\rm st}$ century COE program ``Topological Science and Technology''
and Grants-in-Aid for Scientific Research from the Ministry of Education, Culture, Sports, Science and Technology, Japan.

\end{document}